\documentclass[aps,preprint,superscriptaddress,showpacs]{revtex4}
\usepackage{amsmath,bm}
\usepackage{graphicx}
\usepackage{color}
\begin{document}

\title{Monovacancy and Substitutional Defects in Hexagonal Silicon Nanotubes}
\author{Gunn Kim}
\email[E-mail:\ ]{kimgunn@skku.ac.kr}
\affiliation{BK21 Physics Research Division and Institute of Basic Science, Sungkyunkwan University, Suwon 440-746, Korea}
\author{Suklyun Hong}
\email[E-mail:\ ]{hong@sejong.ac.kr}
\affiliation{Department of Physics and Institute of Fundamental Physics, Sejong University, Seoul, 143-747, Korea}
\date{\today}

\begin{abstract}

We present a first-principle study of geometrical and electronic
structure of hexagonal single-walled silicon nanotubes with a
monovacancy or a substitutional defect. The C, Al or P atoms are
chosen as substitutional impurities. It is found that the defect
such as a monovacancy or a substitutional impurity results in
deformation of the hexagonal single-walled silicon nanotube. In both
cases, a relatively localized unoccupied state near the Fermi level
occurs due to this local deformation. The difference in geometrical
and electronic properties of different substitutional impurities is
discussed.
\end{abstract}
\maketitle

Cubic-diamond bulk silicon is a very well-known semiconducting
material with an energy band gap of 1.2 eV. Unlike the cubic-diamond
silicon, it would be hard to form one-dimensional single-walled
silicon nanotubes (SiNTs) mainly because silicon prefers $sp^3$
bonds to $sp^2$ bonds. However, Bai and his co-workers\cite{Bai}
suggested Si tubes formed by the top-to-top stacking of
square, pentagonal and hexagonal silicon structures and showed using {\it ab
initio} calculations that the pentagonal and hexagonal SiNTs may be
locally stable in vacuum and be metallic. Thus, their work provided
computational evidences for a possible existence of such
one-dimensional silicon nanostructures, although those structures
have not been observed experimentally yet. Therefore, at this point,
it is very valuable to theoretically investigate the stability and
related electronic structure of these silicon nanostructures with
some defects.

In this Letter, we report the modification in geometrical and
electronic structure of hexagonal single-walled silicon nanotubes
with a monovacancy or a substitutional defect. The first-principles
pseudopotential calculations are carried out based on the density
functional theory\cite{Kohn-Sham} within the generalized gradient
approximation for the exchange-correlation functional. The ionic
potential is described with the Vanderbilt-type ultrasoft
pseudopotential.\cite{Vanderbilt} Wave functions are expanded in a
plane wave basis set with an energy cutoff of 21 Ry implemented in
the PWSCF code.\cite{pwscf} As a model system, we choose the
hexagonal SiNT with a monovacancy or a substitutional defect. The C,
Al or P atoms are chosen as substitutional impurities. All chosen
model systems are treated by a supercell with the periodic boundary
condition. The supercell in the lateral direction is as large as
19.2 \AA~and it has 16 times the minimal unit cell (16 $\times$ 2.4
\AA) in the tube axis direction. The structures are relaxed until
the Hellmann-Feynman forces are smaller than 0.05 eV/\AA. The
Brillouin-zone integration is done within the Monkhorst-Pack
scheme\cite{MP} using $1 \times 1 \times 4$ k-point sampling.
To understand the features of charge distribution of our models, we perform the Mulliken population analysis
using a numerical atomic orbital basis set in the OpenMX code\cite{Ozaki1,Ozaki2}
with a kinetic energy cutoff of 150 Ry.
For this Mulliken population analysis, norm-conserving Kleinman-Bylander pseudopotentials\cite{Troullier,Kleinman} are employed.

\begin{figure}[c]
\includegraphics[width=12.5cm,angle=0]{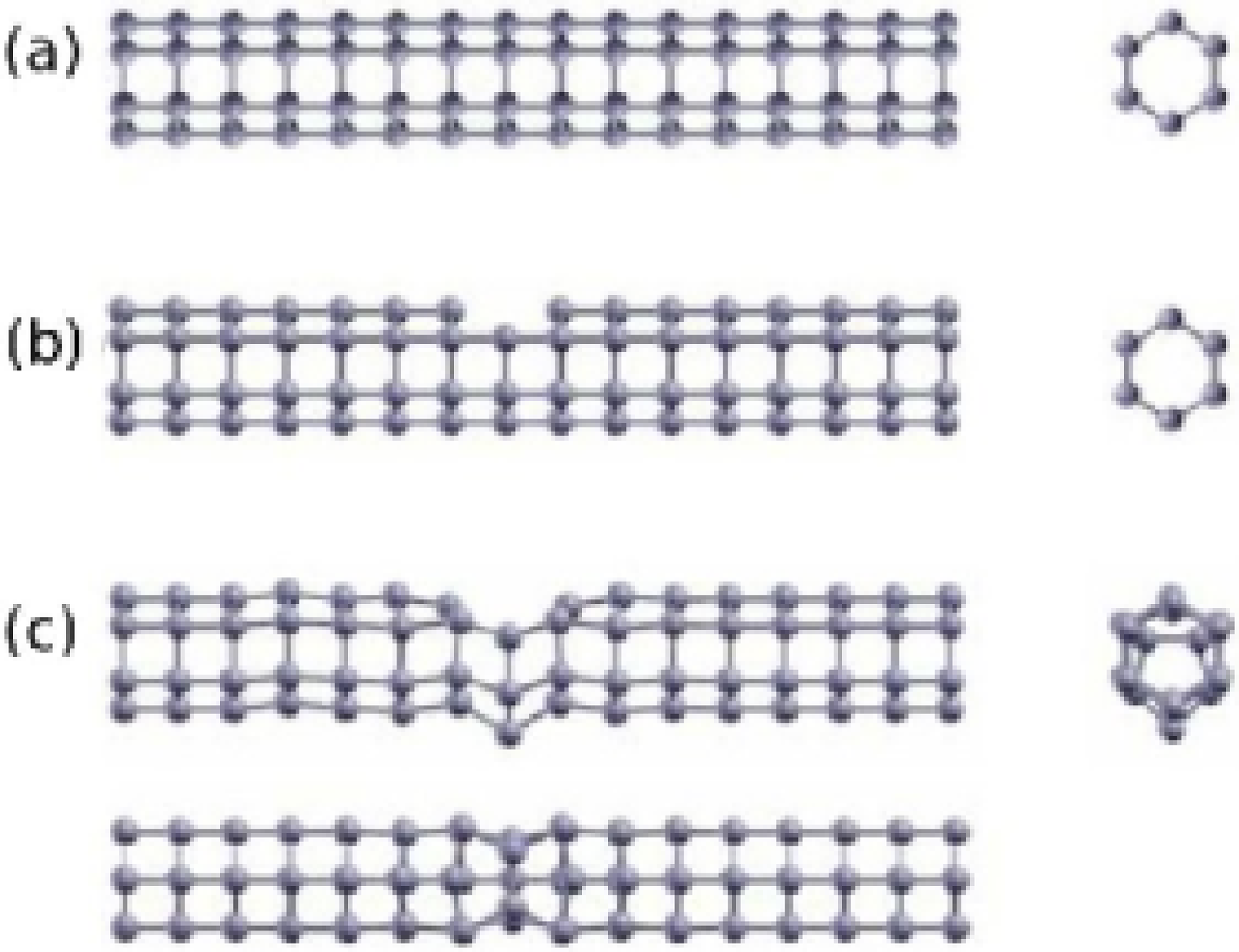}
\caption {Schematic ball-and-stick models of (a) a pristine
hexagonal SiNT, (b) an SiNT with a monovacancy before relaxation,
and (c) an SiNT with a monovacancy after relaxation, respectively.
}\label{model_vac}
\end{figure}

First, we calculate hexagonal SiNTs with or without a monovacancy
shown in Fig.~\ref{model_vac}. If a single Si atom is removed from
the nanotube, two of four Si atoms around the vacancy rebond (bond
length of 2.42 \AA) to form a pentagonal cross section. In the case
of cubic-diamond bulk silicon, the lattice constant is 5.43 \AA,
which corresponds to the Si-Si bond length of 2.35 \AA. Note that a
buckled honeycomb planar structure\cite{Durgun} of silicon has the
average Si-Si bond length of 2.2 \AA. The rebonded two atoms are
perpendicular to the tube axis. Although other two Si atoms have
dangling bonds even after the relaxation, their dangling bond
characters become weak since the two Si atoms make strong bonds with
three adjacent Si atoms (bond length of 2.32$-$2.36 \AA). 
The formation energy ($=E_{perfect}-E_{vacant}-E_{Si~atom}$) of a single vacancy in an SiNT is 16.3 eV (4.07 eV per bond). In the optimized
structure, the Si atom on the opposite side to the monovacancy is
protruded by $\sim$1 \AA~from the surface of the nanotube. The
bonding angle with respect to this protruded Si atom at the
pentagonal defect is 89.6$^\circ$ as shown in Fig. 1(c). Molecular
dynamics (MD) simulations revealed that pentagonal and hexagonal
SiNTs can remain stable.\cite{Bai, Durgun} Our MD simulation result
shows that the SiNT is somewhat distorted but not broken when a
monovacancy is formed at 600 K.

Such locally large distortion may affect the band structure and
charge distribution of the SiNT. In terms of the Mulliken population
analysis, we find an interesting feature. The protruded Si atom on
the opposite side to the vacancy obtains 0.12 $e$ and its adjacent
Si atoms lose electrons: Each of two adjacent Si atoms in the tube
axis direction loses 0.025 $e$, while each of two in the
circumference direction loses 0.035 $e$. On the other hand, the two
rebonded Si atoms lose 0.073 $e$, respectively.

\begin{figure}[c]
\includegraphics[width=12.5cm,angle=0]{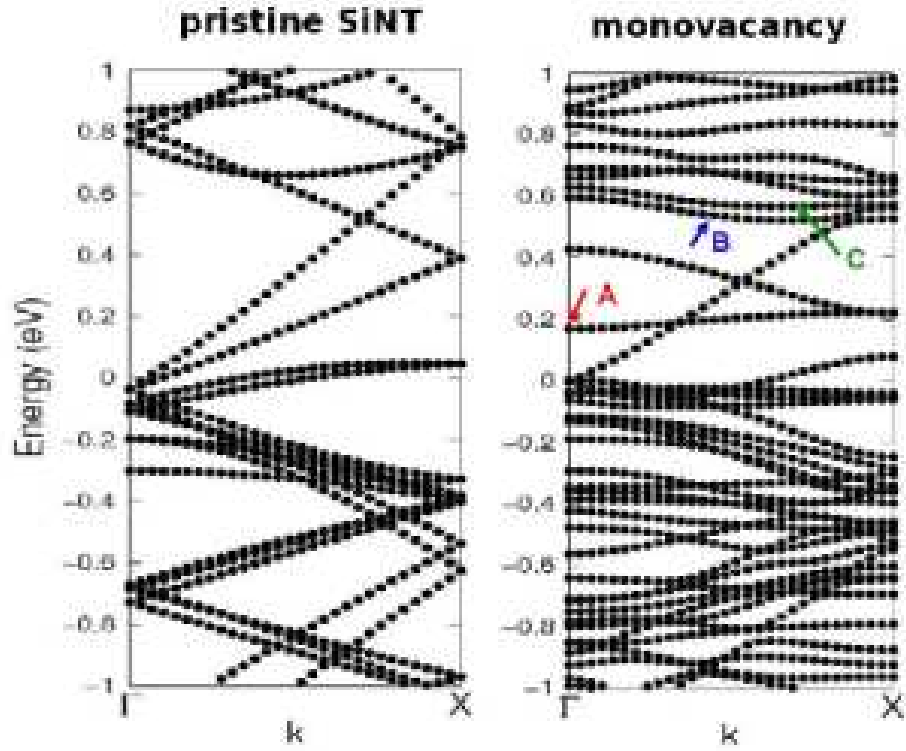}
\caption{Band structures for the SiNT structures corresponding to
(a) a pristine SiNT and (b) an SiNT with a monovacany after
relaxation. States A, B, and C are ULSs. The Fermi level is set to
zero. Localized states occur originating from a monovacancy.
}\label{band_vac}
\end{figure}

\begin{figure}[c]
\includegraphics[width=12.5cm,angle=0]{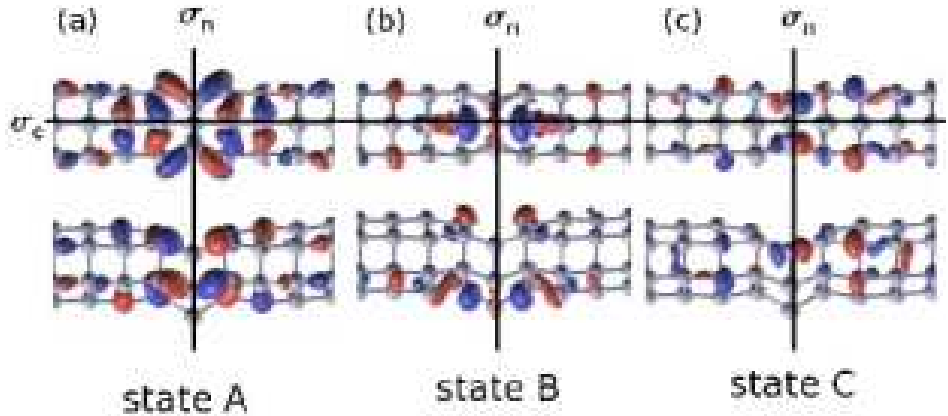}
\caption {Isodensity surface plots of the ULSs at $\Gamma$ that
comes from the distortion due to the presence of vacancy: Top and
side views for a ULS (state A) near $+$0.2 eV due to the local
deformation are shown in (a), and those for two ULSs (states B and
C) near $+$0.6 eV due to the pentagonal defect are in (b) and (c).
The values for the red and blue isodensity surfaces are $\pm$0.02
$e/a_0^3$, where the sign is that of the wave function and
$a_{0}=0.529$ \AA, the Bohr radius. }\label{wf_vac}
\end{figure}

As mentioned above, Fig.~\ref{band_vac}(a) clearly shows that the
hexagonal SiNT has the metallic character.\cite{Bai} When a
monovacancy is formed in the SiNT, the energy band structure is
changed accordingly. There are prominent localized states above the
Fermi level. Especially, a localized state (state A) originating
from the structure distortion gives a flat band around 0.2 eV above
the Fermi level as denoted by a downward arrow in
Fig.~\ref{band_vac}(b). The isodensity surface plot of this
unoccupied localized state (ULS) is presented in
Fig.~\ref{wf_vac}(a). In addition, a pentagon as a topological
defect also gives rise to two ULSs (states B and C) near +0.6 eV as
depicted by upward arrows in Fig.~\ref{band_vac}(b). Interestingly,
Figs.~\ref{wf_vac}(b) and (c) show that the localized states by a
pentagonal defect are split into even and odd parity states with
respect to the mirror planes ($\sigma_c$ and $\sigma_n$) containing
and normal to the nanotube axis, respectively.

\begin{figure}[c]
\includegraphics[width=12.5cm,angle=0]{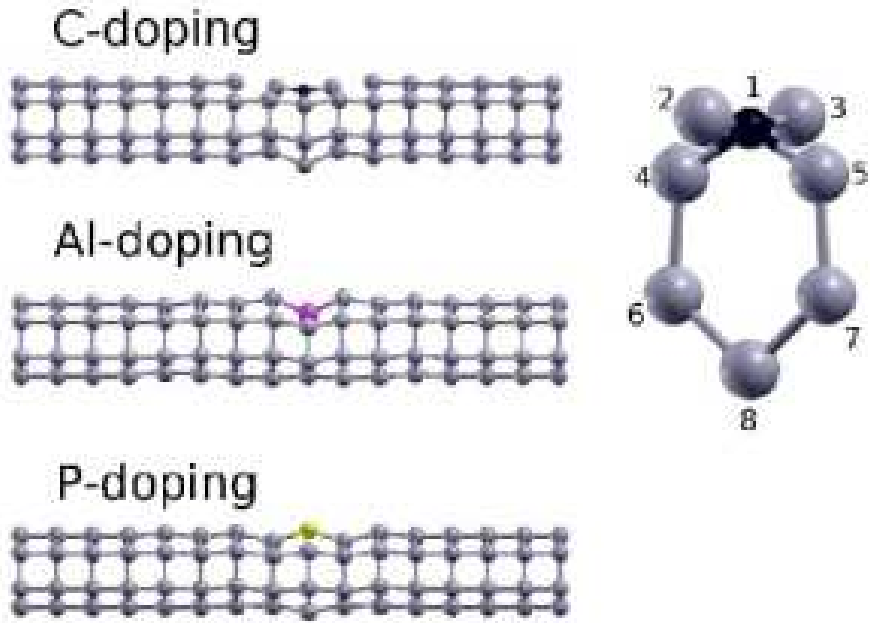}
\caption {Schematic ball-and-stick models with a substitutional
impurity in a hexagonal SiNT. The  grey, black, pink, and yellow
balls represent a silicon, carbon, Al, and P atoms, respectively.
}\label{model_sub}
\end{figure}

Next, we present the geometrical and electronic structure of a
hexagonal SiNT with a carbon, aluminum or phosphorus as a
substitutional impurity, shown in Fig.~\ref{model_sub}. In the case
of the C doping, the C-Si bond length (C1-Si2 and C1-Si3) parallel
to the tube axis is 1.93 \AA~and that (C1-Si4 and C1-Si5)
perpendicular to the axis is 1.92 \AA. The C atom is buckled inward
by $\sim$0.4 \AA~and the Si8 atom on the opposite side to the C
dopant is protruded outward by $\sim$0.6 \AA. Owing to the strain
effect, Si-Si bond breaking occurs near the defect site. The bond
angle between C and adjacent Si atoms, $\angle$Si2C1Si3, parallel to
the tube axis are 165.8$^\circ$~and $\angle$Si4C1Si5 perpendicular
to the axis is 120.6$^\circ$. For the Al doping, the Al-Si bond
lengths parallel and perpendicular to the tube axis are 2.58 and
2.61 \AA, respectively. Here the Al atom is buckled inward by
$\sim$0.7 \AA~and the Si atom on the opposite side to the Al dopant
is buckled inward by $\sim$0.1 \AA. The bond angle between Al and
adjacent Si atoms parallel to the tube axis are 133.1$^\circ$~and
that perpendicular to the axis is 140.3$^\circ$. In contrast, the P
dopant shows an outward buckling by $\sim$0.3 \AA. Such a trend is
shown for boron nitride nanotubes,\cite{Rubio} where nitrogen atoms
are buckled outward and boron atoms are inward, which is associated
with unpaired electrons. For the P doping, the Si atom on the
opposite side to the P dopant is buckled outward by $\sim$0.3 \AA.
The P-Si bond lengths parallel and perpendicular to the tube axis
are 2.50 and 2.30 \AA, respectively. The bond angle between P and
adjacent Si atoms parallel to the tube axis are 145.5$^\circ$~and
that perpendicular to the axis is 102.3$^\circ$. Each impurity (C,
Al or P) atom results in more or less flat bands, i.e., localized
states, (not shown here) about 1.5 eV below or above the Fermi
level. Note that a boron or nitrogen atom doped in the carbon
nanotube also gives the same feature. When substituted, the C atom
donates 0.25 $e$ to the SiNT. The protruded Si atom on the opposite
side to the C atom obtains 0.09 $e$. In contrast, the P atom obtains
0.23 $e$ from the SiNT and the protruded Si atom on the opposite
side to the P atom obtains 0.014 $e$. In the case of the Al
impurity, there is no practical electron transfer between the SiNT
and the Al atom (0.003 $e$). It is thus found that the large
displacement of the Si atom opposite to a substitutional impurity is
associated with the large electron transfer.

\begin{figure}[c]
\includegraphics[width=12.5cm,angle=0]{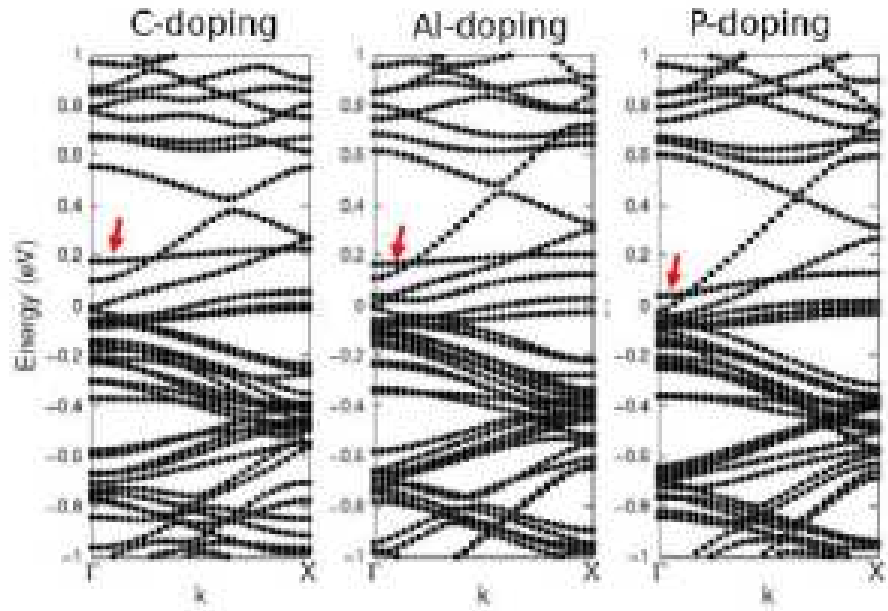}
\caption {Band structures for carbon, aluminum and phosphorus atom
substitutions. Arrows indicate ULS originating from the deformation
near the Fermi level. }\label{band_sub}
\end{figure}

Interestingly, Fig.~\ref{band_sub} shows the relatively flat bands
around 0.1 $-$ 0.2 eV above the Fermi level for the substitution of
C, Al and P, respectively, which are denoted by red arrows. We find
that these localized states originate from the structure distortion,
so the origin of these states is the same as that of the flat band
(denoted by a downward red arrow in Fig.~\ref{band_vac}(c)) of ULS
obtained by the presence of vacancy. Especially, since the
geometrical structure of SiNT with C-doping in Fig.~\ref{model_sub}
is almost the same as that of SiNT with vacancy in
Fig.~\ref{model_vac}(c) except the substituted carbon atom itself,
the isodensity surface plot of the ULS in C-doped SiNT is expected
to be almost the same as that shown in Fig.~\ref{wf_vac}.

In summary, structural and electronic property changes caused by
various vacancy or substitutional defects are studied in the
hexagonal SiNTs using {\it ab initio} pseudopotential calculations.
The C, Al or P atoms are chosen as substitutional impurities. It is
found that the defect such as a monovacancy or a substitutional
impurity results in deformation of the nanotube. The relatively
localized states near the Fermi level shown in both cases with a
monovacancy or an impurity occur due to this local deformation. When
we consider the realistic situation of the randomly distributed
defects, there would be somewhat broadened density of states
originating from almost degenerate localized states.

G.K. is supported by the second BK21 project of Ministry of Education and
S.H. is supported by the grant from the KOSEF through the Center for
Nanotubes and Nanostructured Composites.

\newpage

\newpage

\newpage

\newpage

\newpage

\end{document}